\documentclass[aps,pre,twocolumn,groupedaddress,showpacs,floatfix,superscriptaddress]{revtex4-1}

\usepackage[T1]{fontenc}
\usepackage[utf8]{inputenc}
\usepackage{textcomp} 
\usepackage{graphicx}
\usepackage{amsmath}
\usepackage{amsfonts}
\usepackage{amssymb}
\usepackage{epsfig}
\usepackage{color}
\usepackage{bbm}
\usepackage[hidelinks]{hyperref}

\usepackage{times}

\usepackage{color}

\newcommand{\ep}{\mathcal{E}_p}
\newcommand{\ek}{\mathcal{E}_k}

\begin{document}

\title{Bounding energy growth in friction-less stochastic oscillators}

\author{Micha\l{} Mandrysz}
\email{michal.mandrysz@student.uj.edu.pl} 
\affiliation{Marian Smoluchowski
Institute of Physics, and Mark Kac Center for Complex Systems
Research, Jagiellonian University, ul. St. Łojasiewicza 11,
30--348 Kraków, Poland}

\author{Bart\l{}omiej Dybiec}
\email{bartek@th.if.uj.edu.pl} 
\affiliation{Marian Smoluchowski
Institute of Physics, and Mark Kac Center for Complex Systems
Research, Jagiellonian University, ul. St. Łojasiewicza 11,
30--348 Kraków, Poland}


\begin{abstract}
The paper presents analytical and numerical results on energetics of non-harmonic, undamped, single-well, stochastic oscillators driven by additive Gaussian white noises. Absence of damping and the action of noise are responsible for lack of stationary states in such systems. We explore properties of average kinetic, potential and total energies along with the generalized equipartition relations.
It is  demonstrated that in the friction-less dynamics nonequilibrium stationary states can be produced by stochastic resetting.
For an appropriate resetting protocol average energies become bounded.
If the resetting protocol is not characterized by finite variance of renewal time intervals 
stochastic resetting can only slow down the growth of average energies but does not bound them.
Under special conditions regarding frequency of resets, ratios of average energies follow the generalized equipartition relations.
\end{abstract}


\maketitle


\section{Introduction\label{sec:introduction}}

Energetic properties of stochastic dynamical systems \cite{sekimoto2010stochastic} are determined by the interplay between random forces (fluctuations) and damping (dissipation) \cite{gardiner2009,seifert2012stochastic}.
For damped motions in single-well potentials perturbed by Gaussian white noise stationary states always exist \cite{horsthemke1984,risken1996fokker}.
They are given by the Boltzmann--Gibbs (BG) distribution, which is characterized by finite average energies determined by the system temperature. BG distribution is an example of elliptical distribution because its isolines correspond to constant energy curves, which for the harmonic potential are given by ellipses.
In systems with BG stationary state, ratios of average energies follow generalized equipartition relations \cite{reichl1998,reif2009,mallick2002anomalous}.
Here, we study friction-less dynamics in general single-well potentials using
stochastic \cite{horsthemke1984,hanggi1990,gammaitoni2009} and analytical methods \cite{tome2016stochastic}.
Due to absence of damping, pumping of the energy by noise is not counterbalanced by dissipation.
Therefore, average energies continuously grow in an unbounded manner.
Such unbounded growth of average energies is also responsible for absence of stationary states in friction-less stochastic oscillators.

In order to exclude unlimited energy growth in friction-less dynamics, we suggest a mechanism of bounding the system's energy which is based on the stochastic resetting \cite{evans2019stochastic,evans2011diffusion}.
Stochastic resetting, mainly of the position, is especially important in diffusion processes \cite{evans2011diffusion,kusmierz2015optimal,eule2016non},
search processes \cite{gelenbe2010search,gupta2014fluctuating,meylahn2015large} and multiplicative process \cite{manrubia1999stochastic}. 
In the context of current research, important applications of stochastic resetting include first passage time problems \cite{evans2011diffusion,evans2011diffusion-jpa} and diffusion in potential landscapes \cite{pal2015diffusion}.
According to the Sparre-Andersen theorem \cite{sparre1953,sparre1954,chechkin2003b}
for a free stochastic processes driven by symmetric, Markovian noise 
the first passage time density from the real half line follows the universal $t^{-3/2}$ asymptotics.
The heavy-tailed asympotics of first passage time density explains why the mean first passage time of a free particle to a given target (point) diverges.
Stochastic resetting, by excluding infinite excursions, can make the mean first passage time finite \cite{evans2011diffusion,evans2011diffusion-jpa}.
Typically, in the inverse single-well potentials there is no stationary state, because there is no mechanism which can suppress escaping of particles to the infinity.
Here again, stochastic resetting, which moves a particle back to a fixed point $x_r$, can produce nonequilibrium stationary states in unstable potentials \cite{pal2015diffusion}.
Produced stationary states are of the nonequilibrium type because resetting moves particles from all other points than $x_r$ and introduces a source of probability at the fixed point $x_r$.

Following lines of investigations utilized in \cite{evans2011diffusion,evans2011diffusion-jpa,pal2015diffusion} we use the mechanism of stochastic resetting to bound unlimited energy growth during friction-less dynamics in single-well potentials.
It is assumed that resets are performed at random time instants, while the time between two consecutive resets are independent, identically distributed random variables following a one sided probability density.
During each reset we set the system energy to zero by resetting its velocity and position.
We show that for an appropriate resetting protocol, i.e., for a fine-tuned renewal time distributions, average energies can saturate at any preselected level.  
Moreover, despite the fact that nonequilibrium stationary states are not of the BG type, we study conditions under which average energies satisfy generalized equipartition relations \cite{mallick2002anomalous}.
If renewal time intervals are characterized by the diverging variance, stochastic resetting is not sufficient to reintroduce stationary states.
In such a case we observe a generic slow down of average energies growth rate. 
Here again, we demonstrate that the generalized equipartition relations can be recovered also in situations when stationary states do not exist.

Within current manuscript we study undamped (friction-less) motion in single-well potential of $x^{2n}$ type under action of one or two Gaussian white noise sources.
We continue research initiated in \cite{mandrysz2018energetics} and continued in \cite{mandrysz2019energetics-pre} with the special attention to generalized equipartition relations \cite{mallick2002anomalous} and mechanism of bounding energy growth based on stochastic resetting \cite{evans2011diffusion,pal2015diffusion}.
The model under study, basic theory and main results are presented in Sec.~\ref{sec:model}.
The paper is closed with Summary (Sec.~\ref{sec:summary}) and supplemented with the  Appendix (App.~\ref{sec:evolution}).

\section{Model and Results\label{sec:model}}

Noise perturbed motion in a symmetric single-well potential
\begin{equation}
	V(x)=k\frac{x^{2n}}{2n}\;\;\;\;(k>0,n>0)
	\label{eq:potential}
\end{equation}
is described by the Langevin \cite{risken1996fokker} equation
\begin{equation}
m\frac{d^2x(t)}{dt^2}=-\gamma m  \frac{dx(t)}{dt} - kx^{2n -1}(t) + \sqrt{2 \gamma k_B T m}\xi(t),
\label{eq:full-langevin}
\end{equation}
where $x(t)$ represents the position, $m$ the particle mass, $T$ the system temperature, $k_B$ the Boltzmann constant and $\gamma$ is a damping coefficient.
In Eq.~(\ref{eq:full-langevin}) $\xi(t)$ stands for the Gaussian white noise (GWN) satisfying
\begin{equation}
\langle \xi(t) \rangle=0
\;\;\;\;\;\mbox{and}\;\;\;\;\;
\langle\xi(t) \xi(s) \rangle= \delta(t-s).
\label{eq:gwn-cor}
\end{equation}
The special case of $n=1$ corresponds to the harmonic oscillator \cite{gitterman2005noisy,gitterman2013noisy}, while $n>1$ to anharmonic setups.
In most general situations, $n$ does not need to be integer, in such a case it is necessary to replace $x$ with $|x|$.

The system evolution is perturbed by the Gaussian white noise, which describes interactions of the oscillator with the thermal bath characterized by the temperature $T$.
Action of noise makes position and velocity random variables. 
The probability density $P(x,v;t)$, which is the probability of finding the system in a state characterized by $(x(t),v(t))$, evolves according to the diffusion (Kramers) equation \cite{kubo1966fluctuation,risken1984}
\small
\begin{equation}
 \partial_t P(x,v;t) =\left[\partial_v\left( \gamma v + \frac{V'(x)}{m}   \right) - v \partial_x  +\gamma \frac{k_B T}{m}\partial^2_v \right]P(x,v;t).
 \label{eq:kk}
\end{equation}
\normalsize
Eq.~(\ref{eq:kk}) has the stationary solution  which exist for any potential $V(x)$, such that $V(x) \to \infty$ as $x\to\pm\infty$. It is of the Boltzmann--Gibbs type
\begin{equation}
 P(x,v) \propto \exp\left[ - \frac{1}{k_B T} \left(  \frac{m v^2}{2} + {V(x)} \right) \right].
 \label{eq:st}
\end{equation}
The exponent in Eq.~(\ref{eq:st}) is the total energy $\mathcal{E}$ which is the sum of kinetic $\mathcal{E}_k$ and potential $\mathcal{E}_p$ energies.
The system's total energy $\mathcal{E}=\mathcal{E}_k+\mathcal{E}_p=\frac{1}{2}mv^2+k\frac{x^{2n}}{2n}$ depends on its state $(x(t),v(t))$. Consequently, instantaneous energies, analogous to state variables, are random variables.
Nevertheless, for large $t$ stationary density is reached and  average energies attain constant values.
In the stationary state, the position and the velocity are statistically independent as Eq.~(\ref{eq:st}) factorizes into the product of space dependent and velocity dependent parts.
Moreover, Eq.~(\ref{eq:full-langevin}) assures, that the stochastic harmonic oscillator, corresponding to $n=1$, fulfills the equipartition theorem \cite{kubo1966fluctuation,risken1984}.
Finally, from Eq.~(\ref{eq:st}), for any $n$, one can calculate
\begin{equation}
    \langle \ek \rangle = \iint \frac{1}{2} m v^2P(x,v) dxdv=\frac{1}{2}k_BT,
    \label{eq:avek}
\end{equation}
\begin{equation}
    \langle \ep \rangle = \iint k \frac{x^{2n}}{2n} P(x,v) dxdv=\frac{1}{2n}k_BT
    \label{eq:avep}
\end{equation}
and
\begin{equation}
    \langle \mathcal{E} \rangle = \iint \left[  \frac{1}{2} m v^2 + k \frac{x^{2n}}{2n} \right] P(x,v) dxdv=\frac{n+1}{2n}k_BT.
    \label{eq:ave}
\end{equation}
From Eqs.~(\ref{eq:avep}) -- (\ref{eq:ave}) one obtains
\begin{equation}
    \frac{\langle \ek \rangle}{\langle \mathcal{E} \rangle} = \frac{n}{1+n}
    \label{eq:ek_e_ratio}
\end{equation}
and
\begin{equation}
    \frac{\langle \ep \rangle}{\langle \mathcal{E} \rangle} = \frac{1}{1+n}.
    \label{eq:ep_e_ratio}
\end{equation}
The very same formulas, see Eqs.~(\ref{eq:avek}) and (\ref{eq:avep}), have been also derived in~\cite{mandrysz2020partition} where the undamped, classical, fully deterministic, anharmonic oscillators with $V(x)$ given by Eq.~(\ref{eq:potential}) have been studied.
In other words, undamped classical, conservative, oscillators time averaged kinetic and potential energies are also given by  Eqs.~(\ref{eq:avek}) and (\ref{eq:avep}).
Alternative derivation, for the general nonlinear oscillator with the parametric noise, is presented in Ref.~\cite{mallick2002anomalous}.
Moreover, in accordance with the Virial theorem \cite{goldstein2002classical}
\begin{equation}
    \frac{\langle \ek \rangle}{\langle \ep \rangle}=n.
\end{equation}

In this work, we are interested in modifications of the general model described by Eq.~(\ref{eq:full-langevin}).
More precisely, we are still studying the system described by Eq.~(\ref{eq:full-langevin}) with the $V(x)$ given by Eq.~(\ref{eq:potential}) assuming friction-less dynamics, i.e., dynamics corresponding to $\gamma=0$.
Nevertheless, for the clarity of presentation, we start with the discussion of the full, damped dynamics.
For the purpose of deriving the quantities of interest, Eq.~(\ref{eq:full-langevin}) can be rewritten as a set of two first-order equations
\begin{equation}
\left\{
\begin{array}{ccl}
\frac{dx(t)}{dt} & = & v(t) \\ 
\frac{dv(t)}{dt} & = & -\gamma   v(t) - \omega^2 x^{2n -1}(t) + \sqrt{h_v}\xi(t)
\end{array}
\right.,
\label{eq:set}
\end{equation}
where $\omega^2=k/m$ and $h_v={2 \gamma k_B T}/{m}$.
In Eq.~(\ref{eq:set}) the noise term is present in the second equation only.
Additionally, we assume that also the first equation is subjected to the action of noise \cite{fulinski2000universal}.
Such an extension allow for larger generality than typically studied situation described by Eq.~(\ref{eq:set}).
In particular, two noise sources, allow for study of symmetries of solutions of Eq.~(\ref{eq:two-noises}) especially for the harmonic potential.
Finally, Eq.~(\ref{eq:set}) takes the following form
\begin{equation}
\left\{
\begin{array}{ccl}
\frac{dx(t)}{dt} & = & v(t) + \sqrt{h_x} \xi_x(t) \\
\frac{dv(t)}{dt} & = & -\gamma   v(t) - \omega^2 x^{2n -1}(t) + \sqrt{h_v}\xi_v(t)
\end{array}
\right..
\label{eq:two-noises}
\end{equation}
Evolution of the probability density $P(x,v;t)$ for the system described by Eq.~(\ref{eq:two-noises}) is provided by the diffusion equation, which differs from Eq.~(\ref{eq:kk}) by the presence of the additional $\partial^2_x$ term, see \cite{risken1984}.
Consequently, stationary states for the model described by Eq.~(\ref{eq:two-noises}) cannot be of the BG type.

From Eq.~(\ref{eq:two-noises}) it is possible to derive equations describing time evolution of average kinetic $\langle \ek \rangle$ and potential  $\langle \ep \rangle$ energies, see Appendix~\ref{sec:evolution}.
The time derivatives of average energies are given by
\begin{equation}
 \frac{d}{dt}\langle \ep \rangle= k \langle x^{2n-1} v \rangle + \frac{1}{2}h_x k(2n-1)\langle x^{2n-2} \rangle
\label{eq:epgrowthrate}
\end{equation}
and
\begin{equation}
 \frac{d}{dt}\langle \ek \rangle= -m\gamma \langle v^2 \rangle -m\omega^2 \langle x^{2n-1} v \rangle +\frac{1}{2}mh_v,
 \label{eq:ekgrowthrate}
\end{equation}
while the evolution of the total energy is described by
\begin{eqnarray}
\label{eq:growthrate}
 \frac{d}{dt} \langle \mathcal{E} \rangle & = & \frac{d}{dt} \langle \ep + \ek \rangle \\ \nonumber
 & = & -m\gamma \langle v^2\rangle + \frac{1}{2}h_x k(2n-1)\langle x^{2n-2}\rangle + \frac{1}{2}mh_v. 
\end{eqnarray}

\subsection{Unbounded energy energy growth in the friction-less case\label{sec:unbounded}}

\begin{figure}[htp]
 \centering
 \includegraphics[width=0.9\columnwidth]{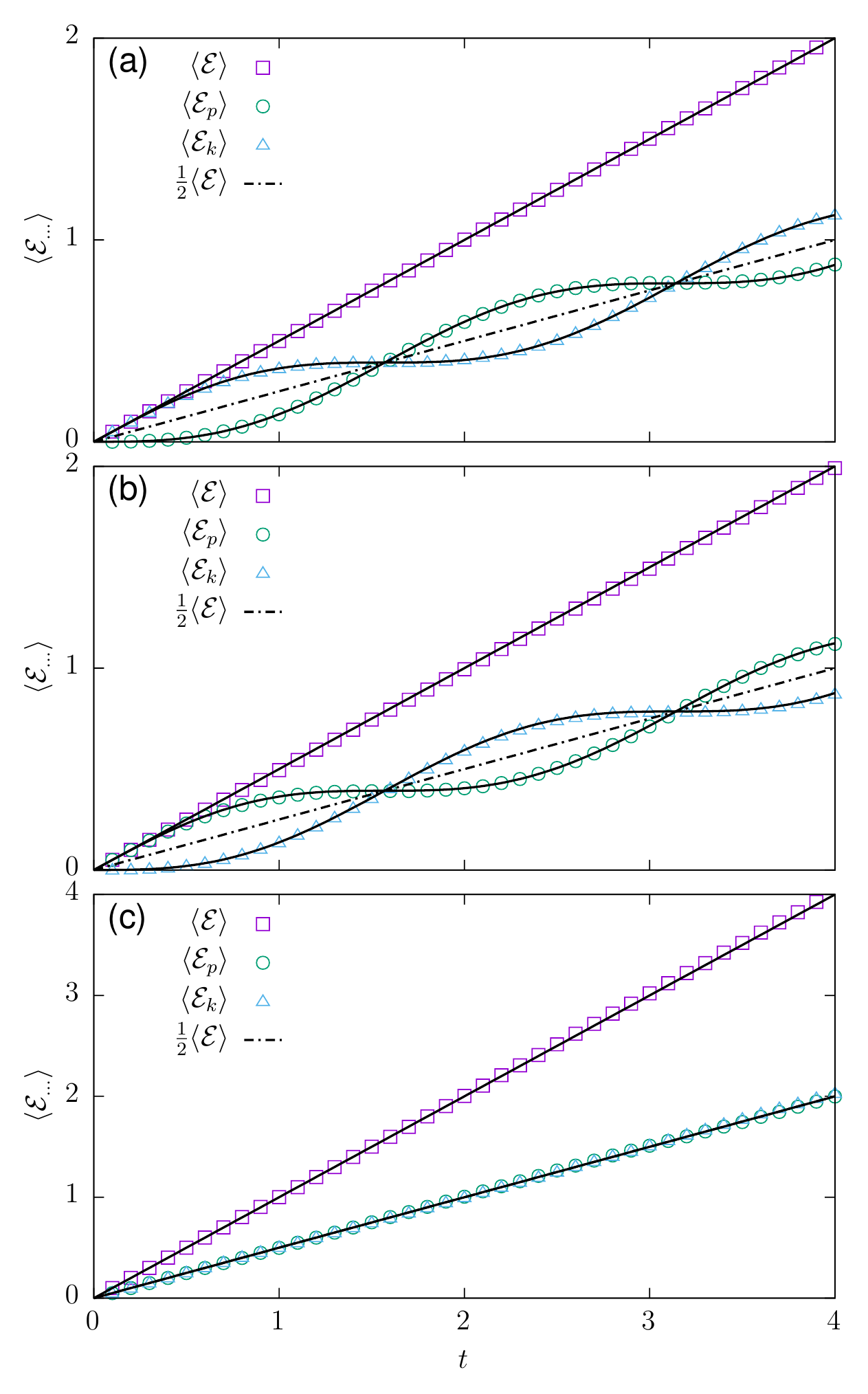}\\
 \caption{Time dependence of average energies for the parabolic potential ($n=1$) with a) $(h_x,h_v)=(0,1)$, b) $(h_x,h_v)=(1,0)$ and c) $(h_x,h_v)=(1,1)$.
 Other parameters $k=1$ and $m=1$.
 Solid lines present theoretical formulas, see Eqs.~(\ref{eq:ek_wn}) -- (\ref{eq:e_wn}) while the dot-dashed line presents $\langle \mathcal{E} (t) \rangle/2$, see Eqs.~(\ref{eq:ek-partition}) and ~(\ref{eq:ep-partition}).
 }
 \label{fig:n2}
\end{figure}

\begin{figure}[htp]
 \centering
 \includegraphics[width=0.9\columnwidth]{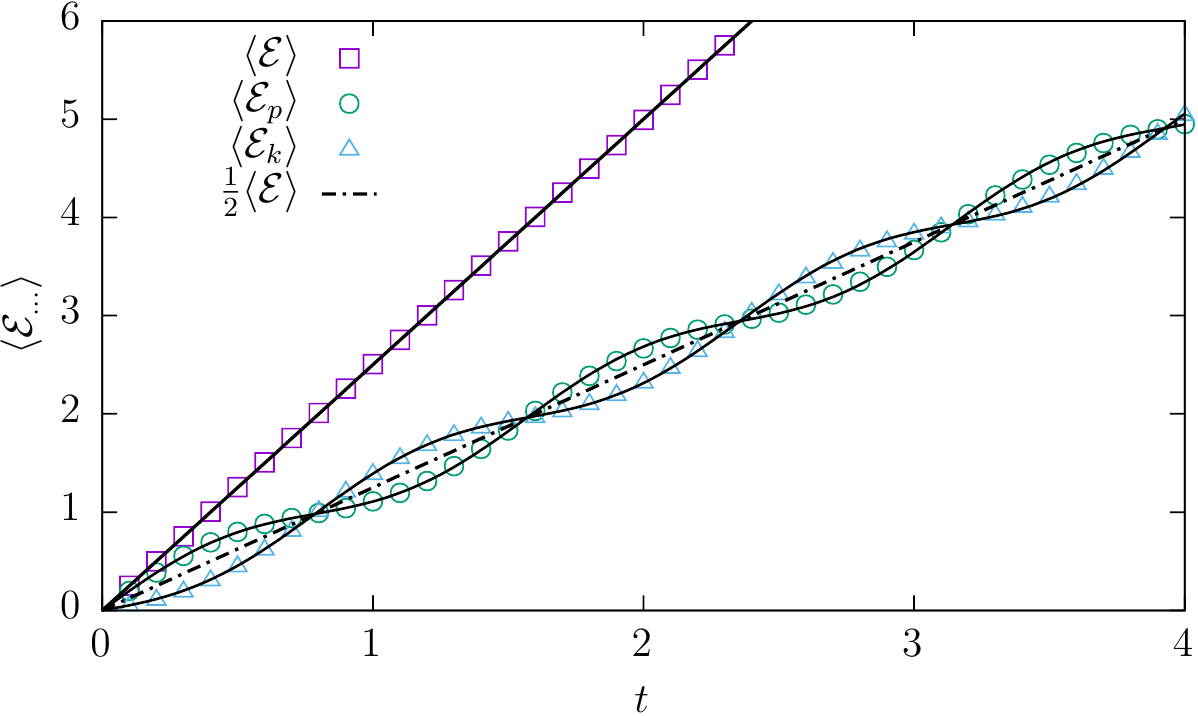}\\
  \caption{The same as in Fig.~\ref{fig:n2} for $k=4$ and $m=1$. Solid lines present theoretical formulas, see Eqs.~(\ref{eq:ek_wn}) -- (\ref{eq:e_wn}).
  The dot-dashed line presents $\langle \mathcal{E}(t) \rangle/2$, see Eqs.~(\ref{eq:ek-partition}) and ~(\ref{eq:ep-partition}).}
  \label{fig:n2omega2}
\end{figure}

\begin{figure}[htp]
 \centering
 \includegraphics[width=0.9\columnwidth]{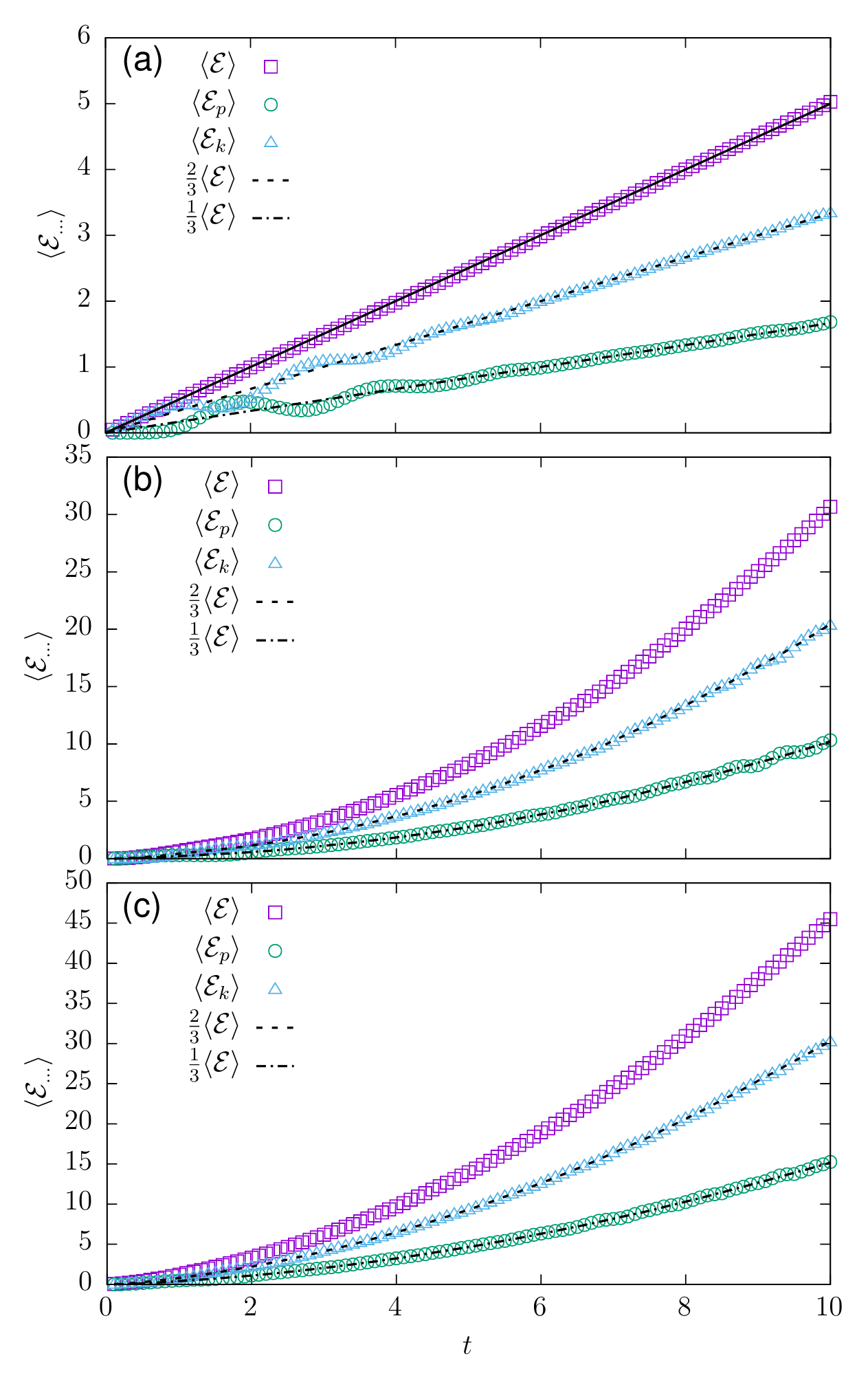}\\
 \caption{Time dependence of average energies for the quartic potential ($n=2$) with a) $(h_x,h_v)=(0,1)$, b) $(h_x,h_v)=(1,0)$ and c) $(h_x,h_v)=(1,1)$.
 Other parameters $k=1$ and $m=1$.
 The solid line in the top panel a) depicts the theoretical linear scaling of the average total energy $\langle \mathcal{E} (t) \rangle$, see Eq.~(\ref{eq:e_wn}).
  Dashed and dot-dashed lines present $\frac{2}{3}\langle \mathcal{E} (t) \rangle$ and $\frac{1}{3}\langle \mathcal{E} (t) \rangle$ scalings, see Eqs.~(\ref{eq:ek-partition}) and ~(\ref{eq:ep-partition}).
 }
 \label{fig:n4}
\end{figure}

For $\gamma>0$ energy growth saturates at stationary value, which is proportional to the system temperature.
In the friction-less case, i.e., for $\gamma=0$ a different situation takes place and average energies grows in unbounded manner \cite{mandrysz2019energetics-pre}.
For $n=1$ the system described by Eq.~(\ref{eq:full-langevin}) or Eq.~(\ref{eq:two-noises}) can be studied analytically \cite{risken1984,mao2007stochastic,tome2015stochastic,czopnik2003frictionless}.
For the parabolic potential these equations are linear and thus standard methods of solving linear differential equations can be applied \cite{risken1984,mao2007stochastic}. 
For $x(0)=0$, $v(0)=0$  following formulas can be derived

\begin{eqnarray}
\label{eq:ek_wn}
\langle \mathcal{E}_k(t) \rangle & = & \frac{1}{2}m \langle v^2 \rangle\\ \nonumber
& = & h_x\frac{2 k t-m \omega \sin(2 \omega t)}{8}+
h_v \frac{2 m \omega t + m \sin (2 \omega t)}{8 \omega } \\ \nonumber
& = & \frac{1}{4}\left[ h_x k + h_v m \right]\times t +\frac{1}{8}\sin(2\omega t)\left[ \frac{h_v m}{\omega} - h_x m \omega \right],
\end{eqnarray}
and
\begin{eqnarray}
\label{eq:ep_wn}
\langle \mathcal{E}_p(t) \rangle & = & \frac{1}{2}m\omega^2 \langle x^2 \rangle
 =  \frac{1}{2}k \langle x^2 \rangle\\ \nonumber
& = & h_x\frac{2 k t+m\omega \sin(2 \omega t)}{8}
+h_v \frac{2m\omega t - m \sin (2 \omega t)}{8 \omega } \\ \nonumber
& = & \frac{1}{4}\left[ h_x k + h_v m \right]\times t +\frac{1}{8}\sin(2\omega t)\left[ h_x m \omega - \frac{h_v m}{\omega}  \right],
\end{eqnarray}
giving rise to
\begin{equation}
\langle \mathcal{E}(t) \rangle =
\langle \ek(t) \rangle+\langle \ep(t) \rangle
=\left[  \frac{h_x k}{2} + \frac{h_v m}{2}  \right] \times t,
\label{eq:e_wn}
\end{equation}
where $\omega=\sqrt{k/m}$.
Presence of the additional noise $\xi_x$ in the first line of Eq.~(\ref{eq:two-noises}) increase the slope of the liner growth of average energy.
Alternatively, Eqs.~(\ref{eq:ek_wn}) and~(\ref{eq:ep_wn}) can also be derived by use of equations for moments, see Eq.~(\ref{eq:n2moments}).
Finally, Eq.~(\ref{eq:e_wn}) can be easily derived from Eq.~(\ref{eq:growthrate}), because for the harmonic oscillator $\langle x^{2n-2} \rangle = \langle 1 \rangle =1$. Such an approach was used in \cite{mandrysz2019energetics-pre} where the $h_x=0$ case was studied.

For the parabolic potential ($n=1$), the linear growth of average energies is clearly visible in Fig.~\ref{fig:n2}, which depicts results for  $\omega=1$ and various values of $h_x$ and $h_v$.
The purely linear growth of average kinetic and potential energies, see Fig.~\ref{fig:n2}c, is recorded due to very special choice of $\omega$.
The special choice of $k$ and $m$ assures that the change in $(h_x,h_v)$ from $(0,1)$ to $(1,0)$ results in a simple exchange of the average kinetic energy with the average potential energy. This symmetry results in the absence of oscillations (equal average energies) at all times, i.e., $\langle \ek (t) \rangle = \langle \ep (t) \rangle$ for $(h_x,h_v)=(1,1)$.
In more general situations this does not occur and hence $\langle \ek (t) \rangle$ and $\langle \ep (t) \rangle$ follow oscillatory growth along the linear trend, see Eqs.~(\ref{eq:ek_wn}) -- (\ref{eq:ep_wn}) and Fig.~\ref{fig:n2omega2}.
Nevertheless, playing with the system's parameters it is possible to control which type of energy dominates at short times. For example, for $h_x=0$, initially $\langle \ek (t) \rangle$ is always larger than $\langle \ep  (t) \rangle$, while for $h_v=0$ average kinetic energy dominates, compare panels a) and c) of Fig.~\ref{fig:n2}.
Moreover, for $n=1$, the average total energy grows linearly regardless of $\omega$, see Eq.~(\ref{eq:e_wn}) and Figs.~\ref{fig:n2} -- \ref{fig:n2omega2}.

Contrary to the harmonic potential, for nonharmonic potentials, the average total energy grows linearly for $h_x=0$ only.
This follows directly from Eq.~(\ref{eq:growthrate}) which gives
\begin{equation}
    \langle \mathcal{E} (t) \rangle = \frac{h_v m}{2}   \times t + \mathcal{E}_0.
    \label{eq:e_nonharmonic}
\end{equation}
Moreover, in~\cite{albeverio1994long} the weak convergence of scaled energy process $ \mathcal{E} (t)/t $ was proven and the scaled stationary probability density $P_{st}(x/t^{1/2n}, v/t^{1/2})$ was derived, see also \cite{mallick2003scaling,mandrysz2019energetics-pre}.

As an exemplary nonharmonic setup, we show results for the quartic ($n=2$) potential. 
In Fig.~\ref{fig:n4}a the predicted linear growth of $\langle \mathcal{E} (t) \rangle$ is observed as $h_x=0$.
In remaining panels average energies grow superlinearly due to presence of $\langle x^{2n-2}\rangle=\langle x^{2}\rangle$ term, see Eq.~(\ref{eq:growthrate}).
From fits, we see that the growth is quadratic.
In all above cases, after disappearing of the transient (usually oscillatory) behavior, generalized equipartition relations hold, see Eqs.~(\ref{eq:ek_e_ratio}) and (\ref{eq:ep_e_ratio}) and Refs.~\cite{mallick2002anomalous,mandrysz2020partition}, i.e., 
\begin{equation}
    \langle \ek \rangle = \frac{n}{1+n}\langle \mathcal{E} \rangle
    \label{eq:ek-partition}
\end{equation}
and
\begin{equation}
    \langle \ep \rangle = \frac{1}{1+n}\langle \mathcal{E} \rangle.
    \label{eq:ep-partition}
\end{equation}
In Fig.~\ref{fig:n4} scalings predicted by Eqs.~(\ref{eq:ek-partition}) and (\ref{eq:ep-partition}) with $n=2$ are depicted with dashed and dot-dashed lines.
These scalings hold not only for the typical noise driven dynamics, i.e., $(h_x,h_v)=(0,1)$, but also for $(h_x,h_v)=(1,1)$ and $(h_x,h_v)=(1,0)$.
%

\begin{figure}[htp]
 \centering
 \includegraphics[width=0.9\columnwidth]{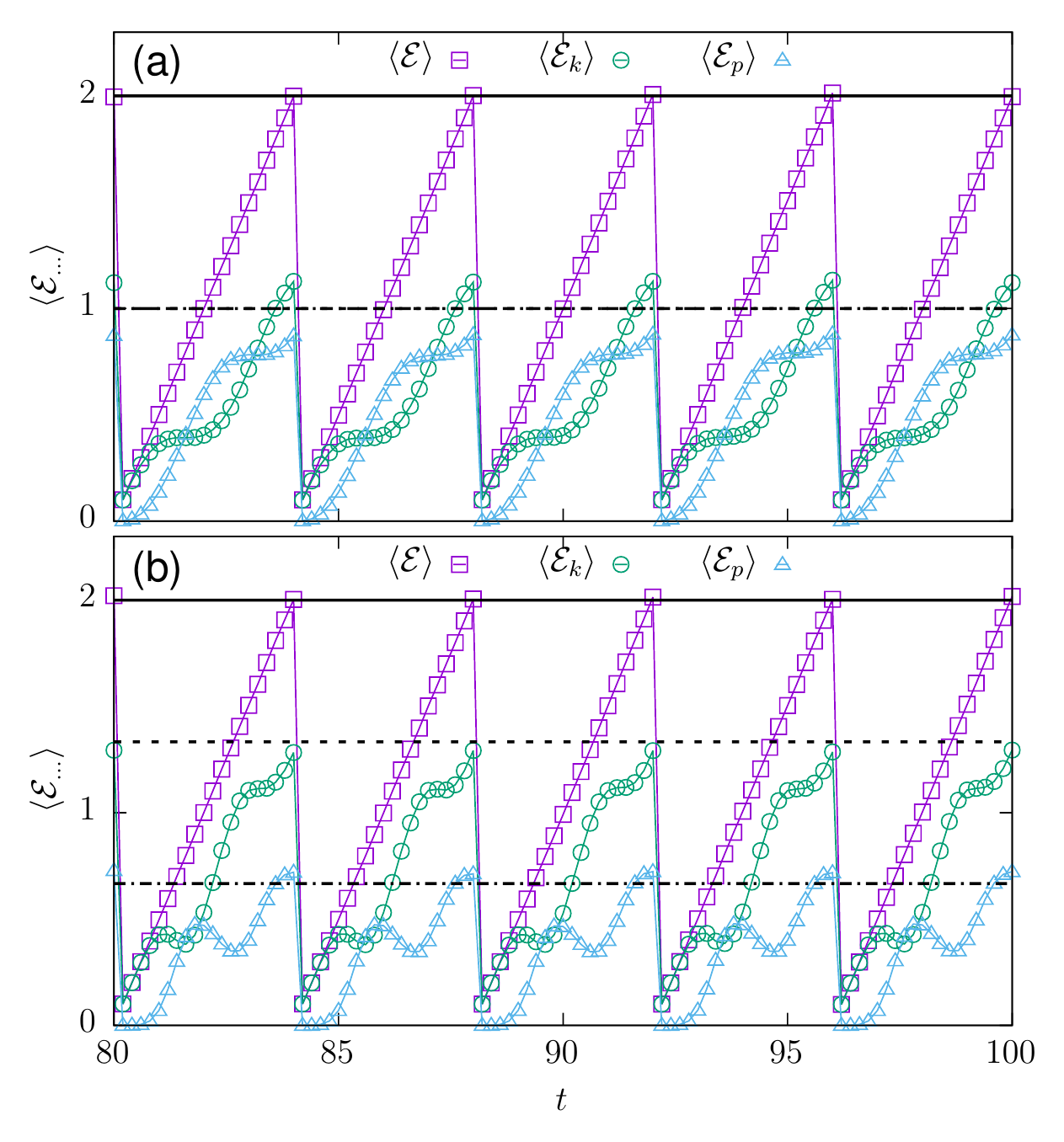}\\
 \caption{Time dependence of average energies for the a) parabolic and b) quartic potentials with
 $(h_x,h_v)=(0,1)$. The reset is performed every $\tau=4$, i.e., $f(\tau)=\delta(\tau-4)$. 
 Other parameters: $k=1$ and $m=1$. 
 Solid, dashed and dot-dashed lines 
 present $\langle \mathcal{E} \rangle_{\mathrm{max}}$ with a)
 $\frac{1}{2}\langle \mathcal{E} \rangle_{\mathrm{max}}$,  $\frac{1}{2}\langle \mathcal{E} \rangle_{\mathrm{max}}$ and 
 $\langle \mathcal{E} \rangle_{\mathrm{max}}$ with b)
 $\frac{2}{3}\langle \mathcal{E} \rangle_{\mathrm{max}}$,  $\frac{1}{3}\langle \mathcal{E} \rangle_{\mathrm{max}}$, see Eqs.~(\ref{eq:ek-partition}) and ~(\ref{eq:ep-partition}).
 }
 \label{fig:n2n4-delta-reset}
\end{figure}

\begin{figure}[htp]
 \centering
 \includegraphics[width=0.9\columnwidth]{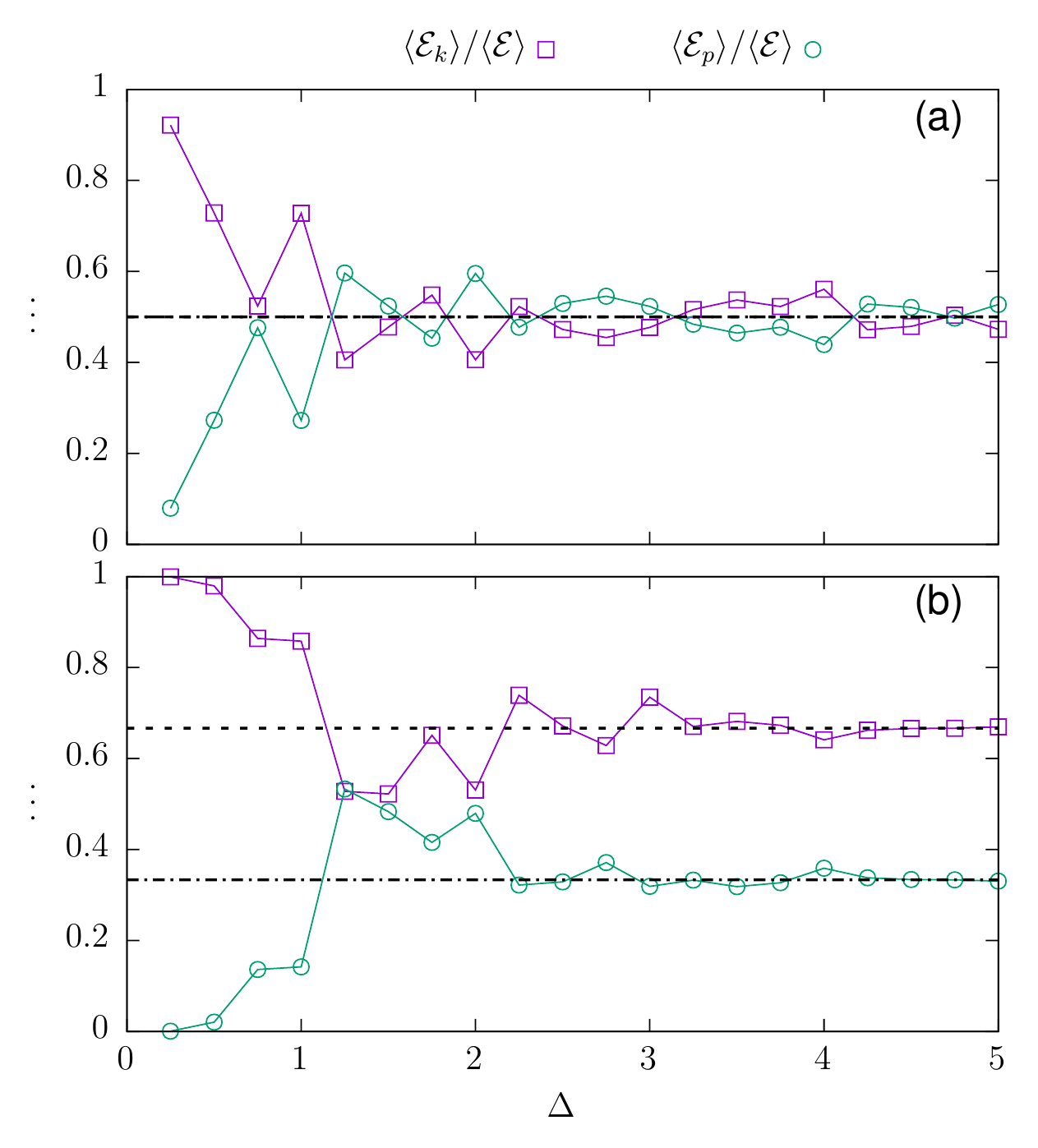}\\
 \caption{Ratio of average energies $\langle \ek \rangle/\langle \mathcal{E} \rangle$ and $\langle \ep \rangle/\langle \mathcal{E} \rangle$ at $t=\Delta$ for the a) parabolic and b) quartic potentials with
 $(h_x,h_v)=(0,1)$. The reset is performed every $\Delta$, i.e., $f(\tau)=\delta(\tau-\Delta)$. 
 Other parameters: $k=1$ and $m=1$. Solid, dashed and dot-dashed lines 
 present theoretical ratios given by Eqs.~(\ref{eq:ek-partition}) and ~(\ref{eq:ep-partition}), i.e., a) $1/2$ and b) $2/3$, $1/3$.
 }
 \label{fig:n2n4-ratio-delta-reset}
\end{figure}

For the parabolic ($n=1$) potential, from Eqs.~(\ref{eq:ek_wn}) -- (\ref{eq:e_wn}) one can calculate asymptotic ($t\to\infty$)  ratio of average energies:
$\langle \ek (\infty) \rangle / \langle\mathcal{E} (\infty) \rangle = \langle \ep (\infty) \rangle/\langle\mathcal{E} (\infty) \rangle =1/2$,
which, for large but finite $t$, agree with predictions of Eqs.~(\ref{eq:ek-partition}) and (\ref{eq:ep-partition}).
Deviations from Eqs.~(\ref{eq:ek-partition}) and (\ref{eq:ep-partition}) are visible for small $t$ because of the periodic addition to the linear trend, see Eqs.~(\ref{eq:ek_wn}) -- (\ref{eq:e_wn}) and Figs.~\ref{fig:n2} -- \ref{fig:n2omega2}.
The very same situation takes place for the quartic ($n=2$) potential, see Fig.~\ref{fig:n4}.
Dashed and dot-dashed lines in Fig.~\ref{fig:n4} present numerically calculated $\langle \mathcal{E} (t) \rangle$ multiplied by factors obtained from Eqs.~(\ref{eq:ek-partition}) and~(\ref{eq:ep-partition}), i.e., $\langle \mathcal{E} (t) \rangle \times n/(1+n)$ and $\langle  \mathcal{E} (t) \rangle \times 1/(1+n)$ with $n=2$.
Expected values of average kinetic and potential energies agree very well with these predictions.


\subsection{Bounding unlimited energy growth}

For $\gamma>0$ the total energy for the model described by Eq.~(\ref{eq:two-noises}) is limited.
For $h_x=0$ the model is equivalent to the standard underdamped Langevin equation~(\ref{eq:full-langevin}) and the stationary density is given by the BG distribution~(\ref{eq:st}).
For $h_x>0$ the stationary density still exist what is clearly visible from computer simulations (results not shown).
Alternatively, one can use equation for moments $\langle x^{2n} \rangle$, $\langle v^2 \rangle$, \dots. For instance, for $n=1$ one has
\begin{equation}
\left\{
\begin{array}{lcl}
\frac{d}{dt} \langle x^2 \rangle & = & 2 \langle xv \rangle +h_x\\
\frac{d}{dt} \langle v^2 \rangle & = & -2\gamma \langle v^2 \rangle -2\omega^2 \langle x v \rangle +h_v\\
\frac{d}{dt} \langle xv \rangle & = &  \langle v^2 \rangle -\gamma \langle xv \rangle -\omega^2 \langle x^2 \rangle\\ 
\end{array}
 \right.,
 \label{eq:n2moments}
\end{equation}
from which Eqs.~(\ref{eq:ek_wn}) and~(\ref{eq:ep_wn}) can also be derived.
Moreover, from Eq.~(\ref{eq:n2moments}) stationary values of moments  $\langle x^2 \rangle,\langle v^2 \rangle$ and $\langle xv \rangle$ can be calculated
\begin{equation}
\left\{
\begin{array}{lcl}
 \langle x^2 \rangle_\infty & = & \frac{h_x (\gamma^2 + \omega^2)+h_v}{2\gamma \omega^2}\\
 \langle v^2 \rangle_\infty & = & \frac{h_x  \omega^2+h_v}{2\gamma}\\
 \langle xv \rangle_\infty & = &  -\frac{h_x}{2}\\ 
\end{array}
 \right..
 \label{eq:n1moments}
\end{equation}
The set of equations for $V(x) \propto x^{2n}$ with $n>1$ is more complicated than for $n=1$.
The increasing complexity is because of higher order terms, e.g., $\langle x^{2n} \rangle$ and $\langle x^{2n-1}v \rangle$, see Eqs.~(\ref{eq:epgrowthrate})  and~(\ref{eq:ekgrowthrate}).
For instance, the formula for the time  derivative of $\langle x^{2n-1}v \rangle$ takes the following form
\begin{eqnarray}
\frac{d}{dt}    \langle x^{2n-1}v \rangle & = &
(2n-1) \langle x^{2n-2} v^2 \rangle \\ \nonumber
& + &  \frac{1}{2}h_x(2n-1)(2n-2) \langle x^{2n-3} v \rangle \\ \nonumber
& - & \gamma \langle x^{2n-1}v \rangle
-\omega^2 \langle x^{4n-2} \rangle,
\end{eqnarray}
because
$
d(x^{2n-1}v)=(2n-1)x^{2n-2}v dx+ \frac{1}{2}(2n-1)(2n-2)x^{2n-3}v(dx)^2+x^{2n-1}dv
$
and Eqs.~(\ref{eq:dx}) and~(\ref{eq:dv}).
Consequently, subsequent equations have to be introduced making the system of equations infinite. 
This should be contrasted with numerically estimated the total average energy $\langle \mathcal{E} (t) \rangle$ and moments $\langle v^2 \rangle \propto \langle \ek \rangle$,  $\langle x^{2n} \rangle \propto \langle \ep \rangle$   which for the friction-less dynamics in the quartic potential, i.e., $n=2$, grow linearly or  quadratically in time, see Fig.~\ref{fig:n4}.

In the damped case, i.e., $\gamma>0$,
the linear restoring force corresponding to $n=1$, see Eqs.~(\ref{eq:n2moments}) -- (\ref{eq:n1moments}), is sufficient to assure existence of stationary state.
More precisely, velocity is bounded due to damping, while position is constrained by the restoring force.
For steeper potentials than parabolic, i.e., $n>1$, the restoring force is stronger making the system localized in smaller fraction of space.
In the limit of $n\to\infty$, the potential well, see Eq.~(\ref{eq:potential}), transforms into infinite rectangular potential well producing the marginal $P(x)$ density uniform.

\begin{figure}[htp]
 \centering
  \includegraphics[width=0.9\columnwidth]{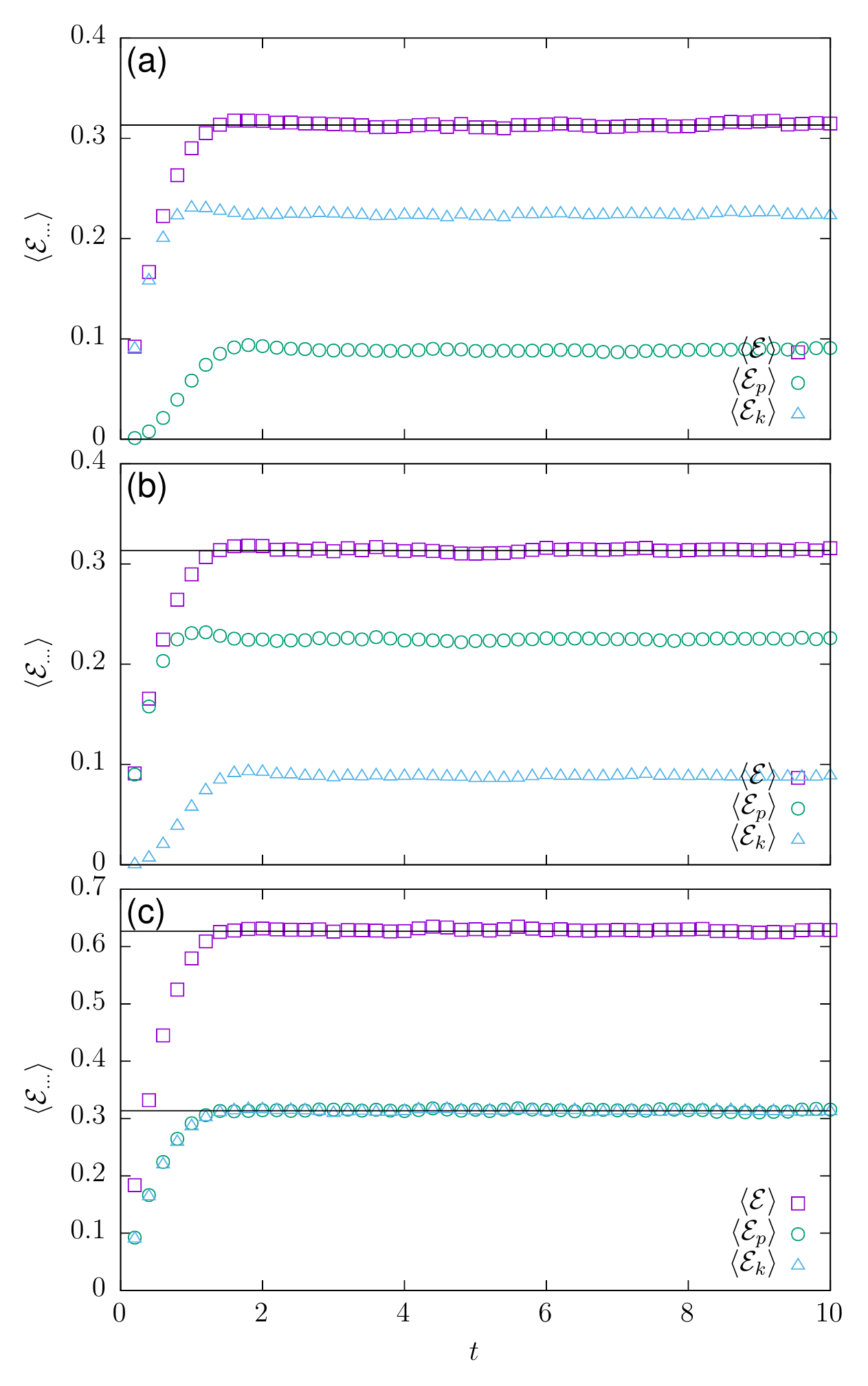} \\
 \caption{Time dependence of average energies for the parabolic ($n=1$)  potential with a) $(h_x,h_v)=(0,1)$, b) $(h_x,h_v)=(1,0)$ and c) $(h_x,h_v)=(1,1)$. Different curves correspond to various types of energies $\langle \mathcal{E} (t) \rangle,\langle\ep (t) \rangle,\langle\ek (t) \rangle$. Other parameters:  $m=1$ and $k=1$.}
 \label{fig:n2-reset}
\end{figure}

\begin{figure}[htp]
 \centering
  \includegraphics[width=0.9\columnwidth]{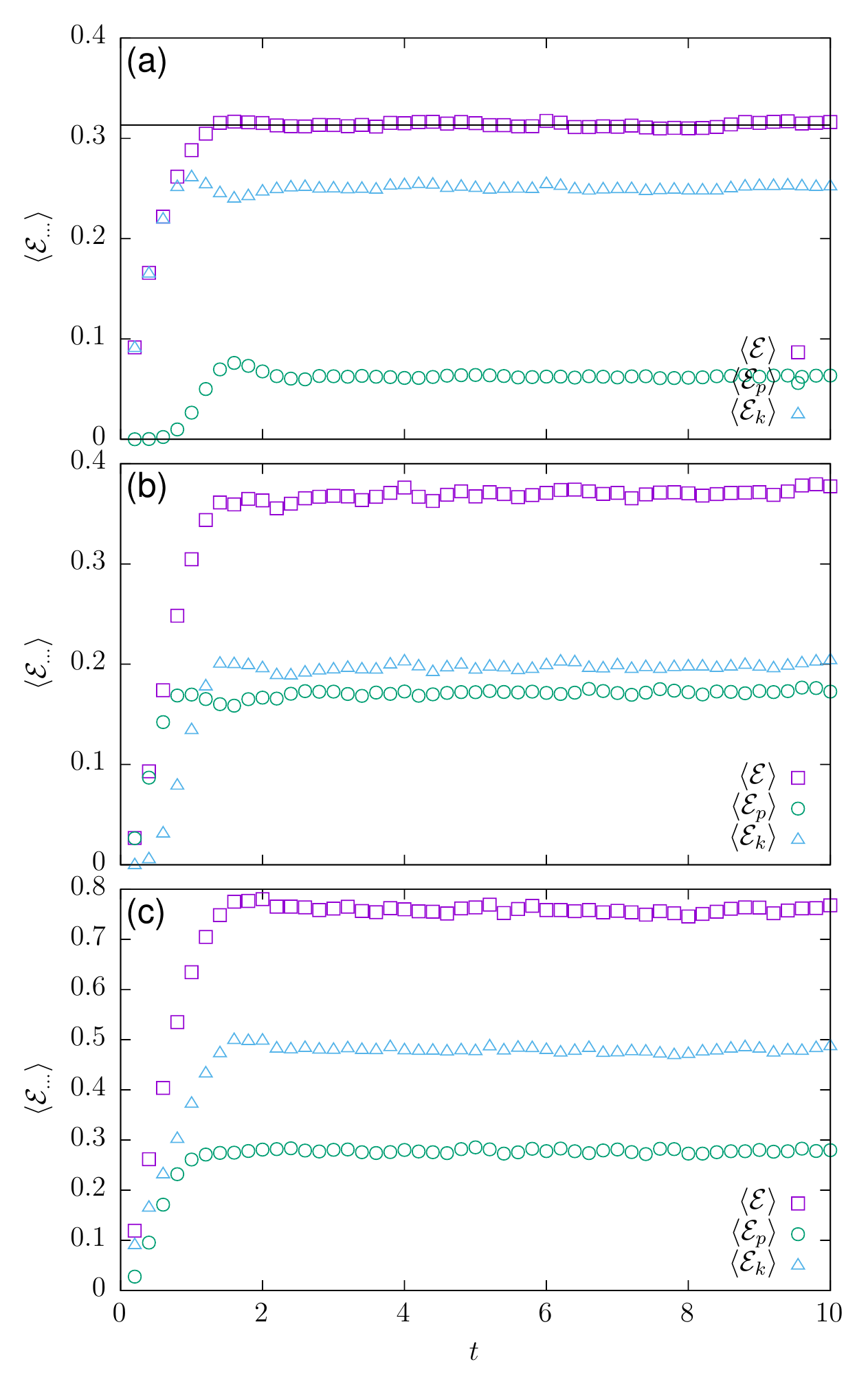}\\
 \caption{The same as in Fig.~\ref{fig:n2-reset} for the quartic ($n=2$) potential.}
 \label{fig:n4-reset}
\end{figure}

\begin{figure}[htp]
   \includegraphics[width=1.0\columnwidth,trim={0.2cm 0.5cm 0.2cm 0.2cm},clip]{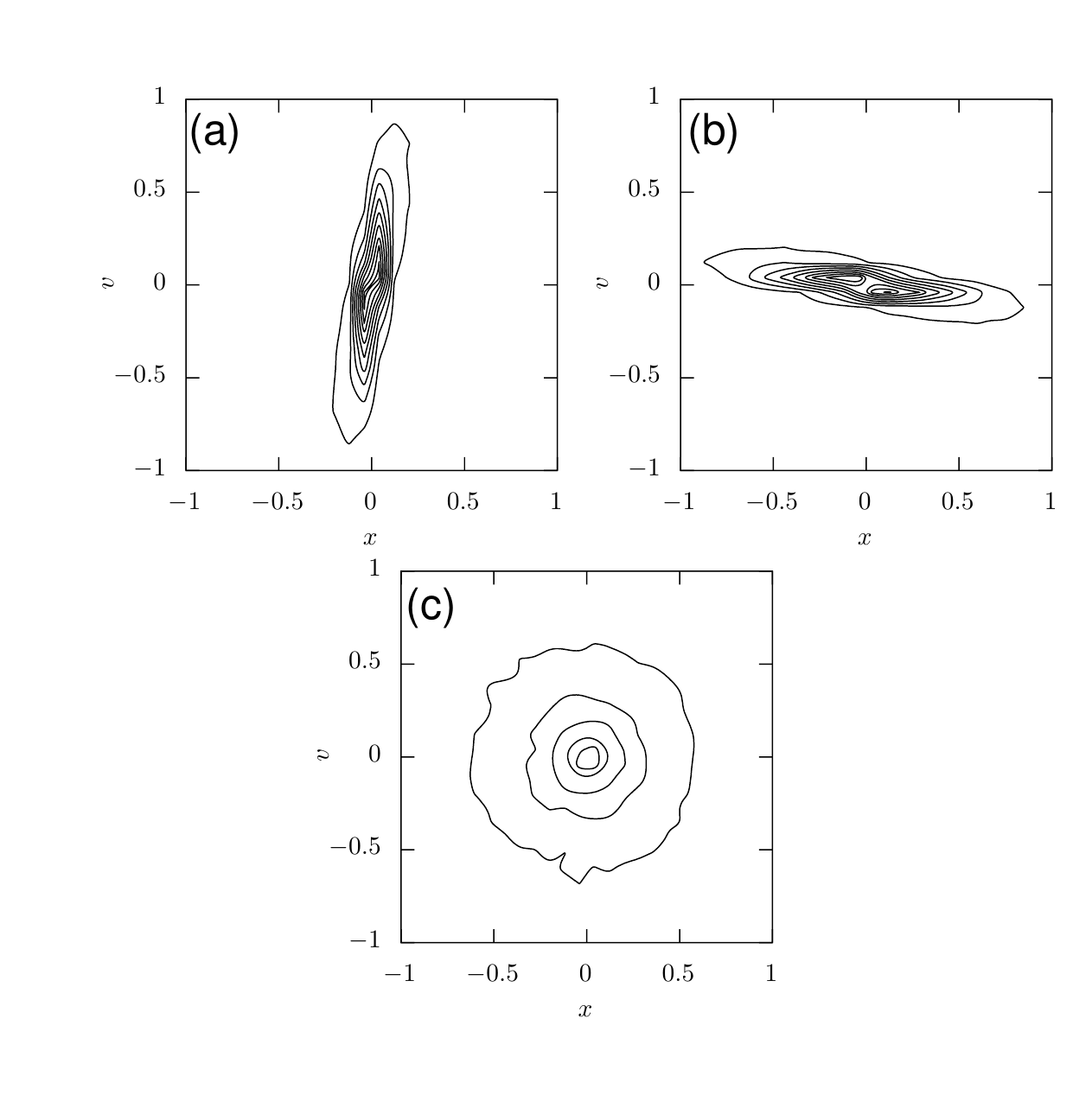} \\
  \caption{Nonequilibrium stationary densities $P(x,v)$ for the parabolic ($n=1$)  potential with a) $(h_x,h_v)=(0,1)$, b) $(h_x,h_v)=(1,0)$ and c) $(h_x,h_v)=(1,1)$. Other parameters:  $m=1$ and $k=1$.}
 \label{fig:n2map}
\end{figure}

\begin{figure}[htp]
 \centering
  \includegraphics[width=0.9\columnwidth]{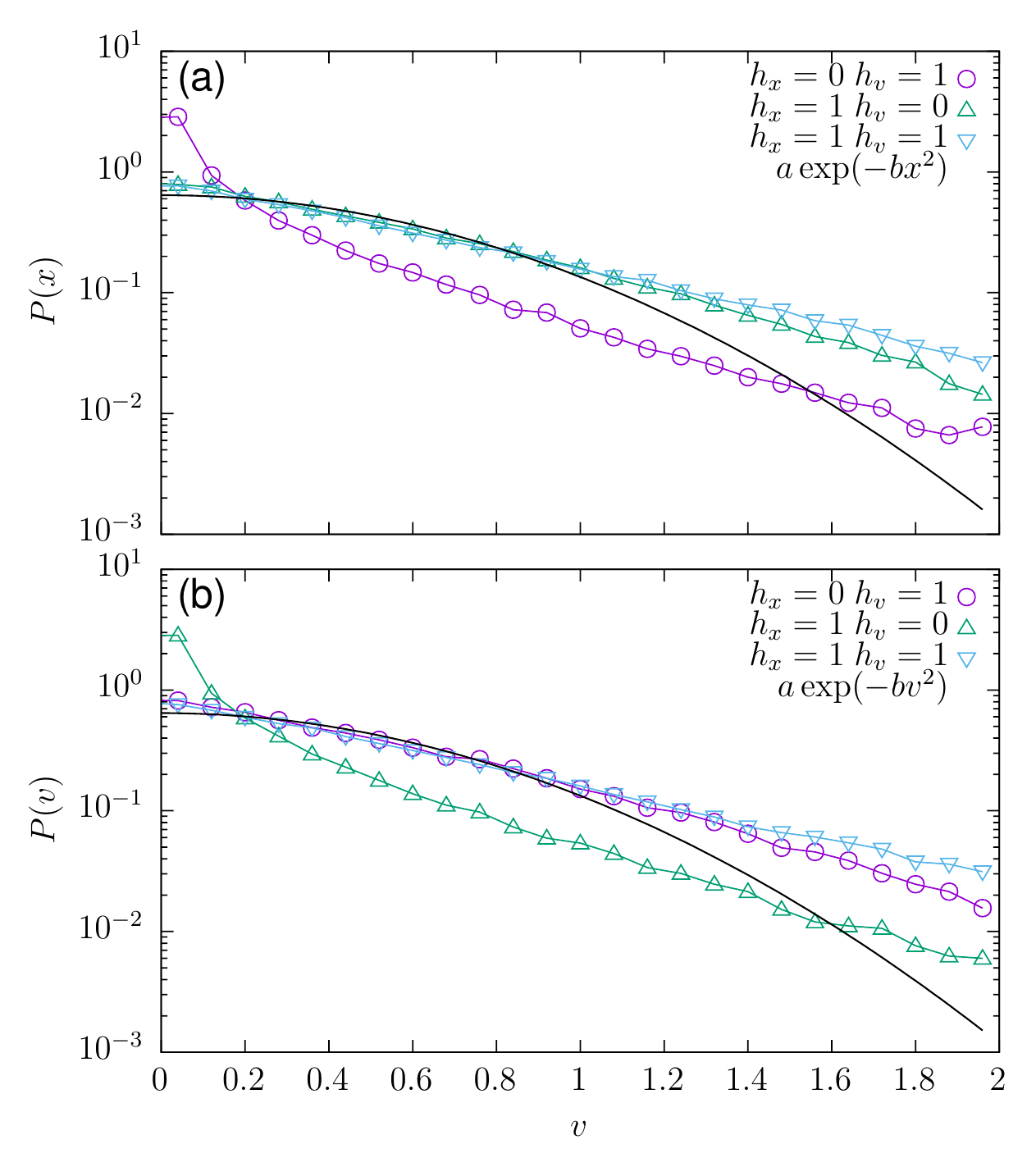} \\

  \caption{Marginal nonequilibrium stationary densities a) $P(x)$ and b) $P(v)$ for the parabolic ($n=1$) potential. Different curves correspond to different values of $h_x$ and $h_v$.
  Solid lines represent $a_x \exp(-b_x x^2)$ and $a_v \exp(-b_v v^2)$ fits.
  Other parameters:  $m=1$ and $k=1$.}
 \label{fig:n2marginal}
\end{figure}

\begin{figure}[htp]
 \centering
  \includegraphics[width=0.9\columnwidth]{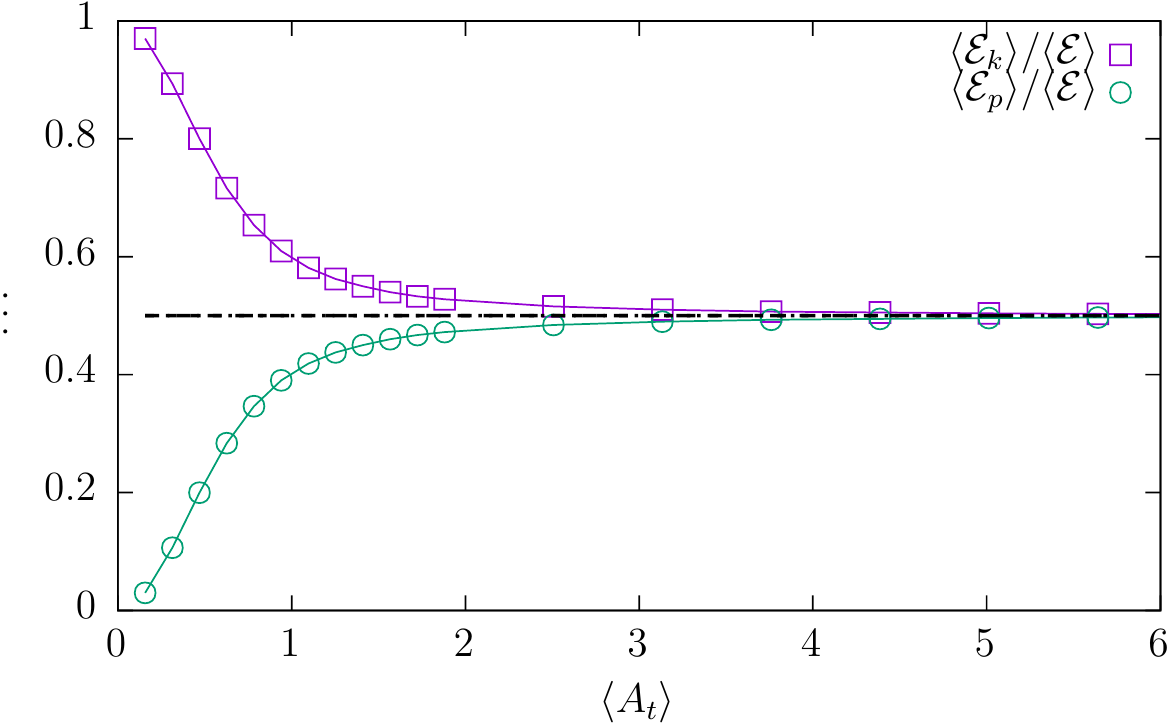} \\
 \caption{Ratio of average energies for the parabolic potential with $(h_x,h_v)=(0,1)$ for the parabolic ($n=1$) potential. Different curves correspond to  $\langle\ep\rangle/\langle \mathcal{E}\rangle$ and $\langle\ek\rangle/\langle \mathcal{E}\rangle$. 
 Dashed and dot-dashed lines depict theoretical ratios predicted by  Eqs.~(\ref{eq:ek-partition}) and ~(\ref{eq:ep-partition}).
 Other parameters:  $m=1$ and $k=1$.}
 \label{fig:n2-reset-sr}
\end{figure}

\begin{figure}[htp]
 \centering
  \includegraphics[width=0.9\columnwidth]{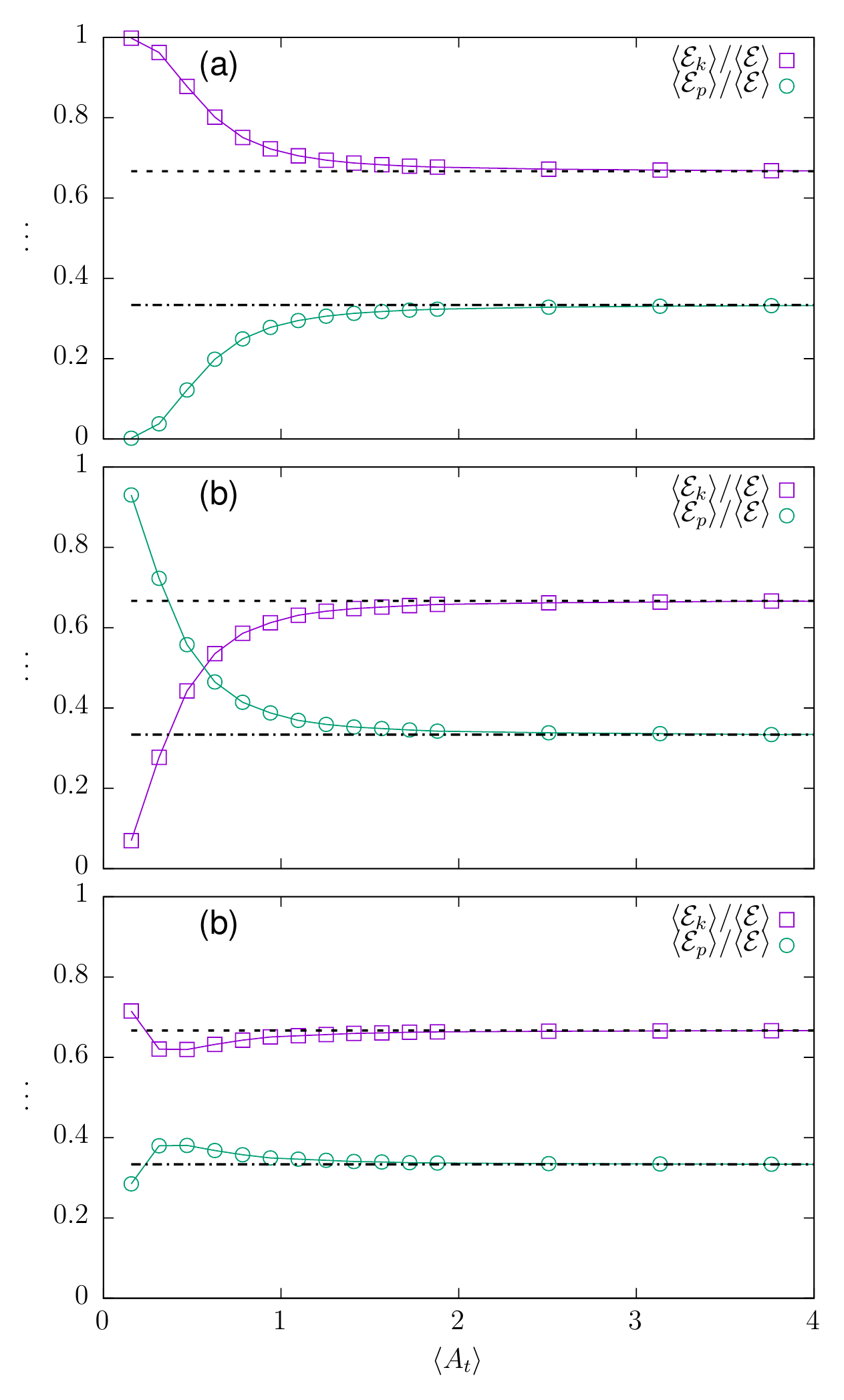} \\
 \caption{The same as in Fig.~\ref{fig:n2-reset-sr} for the quartic ($n=2$) potential with a) $(h_x,h_v)=(0,1)$, b) $(h_x,h_v)=(1,0)$ and c) $(h_x,h_v)=(1,1)$.}
 \label{fig:n4-reset-sr}
\end{figure}

Contrary to the damped dynamics, 
in the friction-less case there is no stationary state for the model described by Eqs.~(\ref{eq:full-langevin}), (\ref{eq:set}) or (\ref{eq:two-noises}).
Lack of damping allows for unbounded energy growth.
Nevertheless, it is possible to introduce other mechanisms resulting in bounding of system's energy.
In order to stop energy growth for $\gamma=0$, we consider stochastic resetting \cite{evans2019stochastic,evans2011diffusion} at random time instants. 
We consider the scenario in which  both velocity and position are simultaneously  reset, i.e., $v\to 0$ and $x\to 0$.
We assume that the resetting is associated with the distribution $f(\tau)$ providing renewal time intervals $\tau$s ($\tau>0$) between resets \cite{breuer2005introduction}. Renewal intervals are independent, identically distributed random variables following $f(\tau)$ distribution.

In the scenario  where both velocity and position are reset the energy is fully determined by the survival time (age) $A_t=t-t_R$, i.e., time measured since the last reset $t_R$.
The sequence of reset times $t_R^{(n)}$ is determined by $f(\tau)$, i.e.,
\begin{equation}
t_R^{(n)}=\sum_{i=1}^n \tau_i ,
\end{equation}
where $\tau_i$ is the sequence of renewal intervals.
As an introductory example, we start with $f(\tau)=\delta(\tau-\Delta)$, where $\delta(\dots)$ is the Dirac's delta. Typical growth of energies $\langle \mathcal{E} (t) \rangle$, $\langle \ep (t) \rangle$ and $\langle \ek (t) \rangle$, interrupted by regular resets occurring every time interval $\Delta$, can be expected.
Indeed, this type of behavior is visible in Fig.~\ref{fig:n2n4-delta-reset} for the parabolic and quartic potentials with $(h_x,h_v)=(0,1)$.
The very similar behavior is recorded for $(h_x,h_v)=(1,0)$ and $(h_x,h_v)=(1,1)$ (results not shown).
Solid lines in Fig.~\ref{fig:n2n4-delta-reset} present maximal total energies while dashed and dot-dashed lines present maximal total energies multiplied by the prefactors from Eqs.~(\ref{eq:ek-partition}) and~(\ref{eq:ep-partition}), which should give maximal average kinetic and potential energies respectively.
As it is visible in Fig.~\ref{fig:n2n4-delta-reset} average kinetic and potential energies not necessarily correspond to the given fraction of total energies.
The violation of equipartition relations in the case of deterministic resets is natural -- there was insufficient time for the system to loose the memory of its initial state, i.e., transient oscillations are present in between reset intervals.
In  Fig.~\ref{fig:n2n4-ratio-delta-reset} we present ratios of average energies as a function of the time interval $\Delta$ between two consecutive resets.
Fig.~\ref{fig:n2n4-ratio-delta-reset} clearly indicates the return to general equipartition (scalings predicted by Eqs.~(\ref{eq:ek-partition}) and~(\ref{eq:ep-partition})) in the case of large $\Delta$.
Consequently, if $\Delta$ is too small general equipartition cannot be satisfied.

The more interesting situation is when resets are performed at random time instants.
In such a case, in order to calculate the average energy in addition to averaging over noise realizations it is necessary to perform additional averaging over time (age) which passed since the last reset.
The rate of average energy growth is easily traceable analytically in situations when in the absence of resetting average energy grows linearly in time.
As we have shown in Sec.~\ref{sec:unbounded} such situation takes place for $\gamma=0$ with $n=1$, see Eq.~(\ref{eq:e_wn}) or any $n$ with $h_x=0$, i.e.,
\begin{equation}
 \langle \mathcal{E}(t) \rangle  = \Lambda t
\end{equation}
with
\begin{equation}
\label{eq:a}
\Lambda =\left\{
\begin{array}{lcl}
\frac{h_x k+h_v m}{2}  &  \mbox{for} & n=1\\
 \frac{h_v m}{2} & \mbox{for} &  n>1, h_x=0.
\end{array}
\right.
\end{equation}
The value of average energy is then determined by properties of renewal intervals, which are times between two consecutive resets.
Here, we assume that renewal intervals $\tau_i$ are independent and identically distributed random variables following the $F(\tau)$ distribution ($f(\tau)=\frac{dF(\tau)}{d\tau}$).
After transient period, the average energy does not depend on time if the average time since the last reset (age) is finite. 
The average time since the last reset $\langle A_t \rangle$ exist if the variance of renewal intervals $\tau$ is finite, i.e., $\sigma^2(\tau)=\langle \tau^2 \rangle - \langle \tau \rangle^2 < \infty $.
In such a case, from the renewal theory \cite{breuer2005introduction}, we have the following formula for $\langle A_t \rangle$
\begin{equation}
 \langle A_t \rangle  =  \frac{\langle \tau^2 \rangle}{2 \langle \tau \rangle }, 
\end{equation}
where  $\langle \tau \rangle$ and $\langle \tau^2 \rangle$ stands for moments of the renewal time intervals, e.g.,  $\langle \tau \rangle = \int_0^\infty f(\tau) \tau d\tau$.
The average energy reads 
\begin{equation}
 \langle \mathcal{E}(t) \rangle  =  \Lambda \langle A_t \rangle  = \Lambda \frac{\langle \tau^2 \rangle}{2 \langle \tau \rangle } 
 \label{eq:mean-finite-stv}.
\end{equation}
Figs.~\ref{fig:n2-reset} -- \ref{fig:n4-reset} present time dependence of the average energies for parabolic $n=1$   (Fig.~\ref{fig:n2-reset}) and quartic $n=2$ (Fig.~\ref{fig:n4-reset})  potentials with the half-normal renewal time distribution
\begin{equation}
f(\tau)=\sqrt{\frac{2}{\pi \sigma_r^2}}\exp\left[-\frac{\tau^2}{2 \sigma_r^2}\right] \;\;\;\;\;\ (\tau>0) 
\label{eq:onesidedgauss}
\end{equation}
for which
$\langle \tau \rangle = \sqrt{2/\pi}\sigma_r$, 
$\langle \tau^2 \rangle = \sigma_r^2$,
$\sigma^2(\tau)=(1-2/\pi)\sigma_r^2$
and the most importantly
$\langle A_t \rangle =  \sqrt{\pi/8}   \times \sigma_r$.
Consequently, for $n=1$ or $h_x=0$  we have
\begin{equation}
\langle \mathcal{E} \rangle= \Lambda \langle A_t \rangle= \Lambda \sqrt{\frac{\pi}{8}} \sigma_r,
\label{eq:resetlimit}
\end{equation}
where $\Lambda$ is given by Eq.~(\ref{eq:a}).
The above predictions, are confirmed in Figs.~\ref{fig:n2-reset} and \ref{fig:n4-reset} which present results for $n=1$ and $n=2$ respectively.
For the parabolic potential the agreement between theoretical predictions, see Eq.~(\ref{eq:resetlimit}), and computer simulations is reached in all cases.
For $(h_x,h_v)=(1,0)$ and $(h_x,h_v)=(0,1)$ the average total energies $\langle \mathcal{E} \rangle$ are the same, while for $(h_x,h_v)=(1,1)$ it is two times larger than in former cases.
For the quartic potential, we have the formula for the stationary value of $\langle \mathcal{E} \rangle$ for $h_x=0$  only, see Fig.~\ref{fig:n4-reset}a, because only for $h_x=0$ we know the formula for $\langle \mathcal{E}(t) \rangle$ (see Eq.~\ref{eq:a}).
For the quartic potential with $h_x>0$, see panels b) and c) of Fig.~\ref{fig:n4-reset}, average energies are also bounded.
At this time stationary values are not easily related to $\langle A_t \rangle$  because in the absence of resetting average energies are nonlinear functions of time, see Fig~\ref{fig:n4}.
Analogously like for $n=1$, for $(h_x,h_v)=(1,1)$, due to presence of two noise sources, average energies attain larger values.
In Figs.~\ref{fig:n2-reset} and \ref{fig:n4-reset} average energies partitions differ from predictions of Eqs.~(\ref{eq:ek-partition}) and~(\ref{eq:ep-partition}). The difference, in the case of $n=1$, comes from regular oscillations and relatively small $\langle \tau \rangle$. In the special case of $h_x=h_v=1$ and $m=k=1$ when we had no oscillations (see Fig.~\ref{fig:n2}c) the problem did not appear, however the equipartition was accidental, see below.

The saturation of average energies, see Figs.~\ref{fig:n2-reset} and~\ref{fig:n4-reset}, is connected with the existence of nonequilibrium stationary states.
For the parabolic potential, nonequilibrium stationary densities corresponding to Fig.~\ref{fig:n2-reset}  are depicted in Fig.~\ref{fig:n2map}.
Various panels of Fig.~\ref{fig:n2map} correspond to a) $(h_x,h_v)=(0,1)$, b) $(h_x,h_v)=(1,0)$ and c) $(h_x,h_v)=(1,1)$.
If nonequilibrium stationary densities would be of the BG type, these densities would be constant on the constant energy curves, i.e., they would be elliptical.
In case of $n=1$ with $k=1$ and $m=1$, these ellipses should reduce to circles.
Therefore, for $(h_x,h_v)=(0,1)$ and $(h_x,h_v)=(1,0)$ nonequilibrium stationary densities are clearly not of BG type.
For $(h_x,h_v)=(1,1)$ the nonequilibrium stationary density appear to be spherically symmetric, however this is only one of the properties of the BG distribution. 
To finalize our tests, we present the additional Fig.~\ref{fig:n2marginal} with marginal densities a) $P(x)$ and b) $P(v)$. Fig.~\ref{fig:n2marginal} confirms that nonequilibrium stationary densities differ from BG type also for $(h_x,h_v)=(1,1)$.
Please note, that for $\gamma>0$ and $h_x>0$ the stationary state is not of the BG type.
Finally, we have checked that also in situations when generalized equipartition relations hold, i.e., for $\sigma_r$ large enough, nonequilibrium stationary densities are not of the BG type (results not shown).
In particular, the distributions appear to have elongated tails compared with the BG distribution. 
In all above mentioned cases the nonequilibrium stationary states exist due to rapid decay of the tails of the renewal time distribution (faster than $\tau^{-2}$), see \cite{evans2019stochastic} and \cite{nagar2016diffusion}. 

Violations of the generalized equipartition relations visible in Fig.~\ref{fig:n2-reset} and~\ref{fig:n4-reset} are produced by resetting. 
The parameter $\sigma_r$ controls the mean value and the variance of renewal intervals $\tau$, see Eq.~(\ref{eq:onesidedgauss}).
It is responsible for spreading of $A_t$ and increase in $\langle A_t \rangle$.
In order to reintroduce equipartition relations, the parameter $\sigma_r$ has to be large enough to increase the mean survival time (age) $\langle A_t \rangle$  beyond the transient oscillatory phase.
If we increase $\sigma_r$ and consequently $\langle A_t \rangle$ ratios of average energies become closer to predictions of Eqs.~(\ref{eq:ek-partition}) and~(\ref{eq:ep-partition}), see Figs.~\ref{fig:n2-reset-sr} and~\ref{fig:n4-reset-sr}.
Figs.~\ref{fig:n2-reset-sr} and~\ref{fig:n4-reset-sr} show ratios of average energies for parabolic and quartic potentials as a function of the average age $\langle A_t \rangle$.
These figures suggest that violation of general equipartitions relations is caused by too frequent resets.
Moreover, they indicate that if the average age $\langle A_t \rangle$ is longer than the duration of the transient phase general equipartition relations are recovered.
For $\langle A_t \rangle$ large enough generalized relations, see Eqs.~(\ref{eq:ek-partition}) and~(\ref{eq:ep-partition}), are in tact for all studied setups, i.e., also under action of two noise sources, see Fig.~\ref{fig:n4-reset-sr}.
Therefore, for random resets the situation resembles already discussed problem of ratios of energies at fixed times for equidistant resets, see Fig.~\ref{fig:n2n4-ratio-delta-reset}. 
In Fig.~\ref{fig:n2n4-ratio-delta-reset}  generalized equipartition holds (at reset time points) for large enough $\Delta$.

If the variance $\sigma^2(\tau)$ of the renewal interval diverges the average energy $\langle \mathcal{E} (t) \rangle$ does no longer saturate but it starts to grow, because there is no stationary state in the system \cite{nagar2016diffusion}.
The energy growth cannot be larger than in the without resetting dynamics which limit the overall growth rate.
In particular, for $n=1$ or any $n$  with $h_x=0$ the growth is sublinear, as the linear growth is the limiting growth which is recovered in the absence of resetting, see Eq.~(\ref{eq:growthrate}).
In order to calculate the time dependence of the average energy one needs to know the distribution $g(A_t)$ of the time since the last reset, i.e., the age at time $t$. The age $A_t$ is given by $A_t=t-t_R$ where $t_R$ is the last resetting (renewal) time. 
The age distribution fulfills \cite{breuer2005introduction}
\begin{equation}
 P(A_t\leqslant s ) =
 \left\{
 \begin{array}{ll}
  F(t)-\int_0^{t-s}(1-F(t-y))dR(y) & s<t \\
  1 & s \geqslant t
 \end{array}
 \right.,
\end{equation}
where $F(t)$ is the renewal intervals distribution ($F(t)=\int^t f(s) ds$) and $R(t)=\langle N_t \rangle$ with $N_t:= \max\{ n\in \mathbb{N} : \sum_{i=1}^n \tau_i \leqslant t \}$.
Now, the average energy can be calculated as
\begin{equation}
 \langle \mathcal{E} \rangle = \Lambda \int_0^t g_t(s) s ds,
\end{equation}
where 
\begin{equation}
 g_t(s)=\frac{dP}{ds}= 
 \left\{
 \begin{array}{ll}
  (1-F(s)) R'(t-s) & s<t \\
  (1-F(s)) R'(0) \delta(t-s)  & s=t \\
  0 & s > t
 \end{array}
 \right..
\end{equation}
If $\langle \tau \rangle$ is finite for $t \to \infty $, due to Blackwell's theorem \cite{breuer2005introduction}, we can approximate $R(t)$ by $t/\langle \tau \rangle$ and $R'(t)$ by $\frac{1}{\langle \tau \rangle }$. 
Moreover, for the Pareto density, see Eq.~(\ref{eq:pareto}), characterized by the finite mean we can be calculate $R'(0)$ which is equal to 0 because 
\begin{equation}
\frac{dR}{dt}(t)=\lim_{dt\to 0}\frac{\langle N_{t+dt}\rangle - \langle N_{t}\rangle}{dt}    
\end{equation}
and for $dt<\delta$ we get $N_{0+dt}=0$.
Consequently, we have
\begin{equation}
 \langle \mathcal{E} \rangle = \Lambda \int_0^t g_t(s) s ds=\frac{\Lambda}{\langle \tau \rangle} \int_0^t (1-F(s)) s ds.
\end{equation}
The estimate of $\langle \mathcal{E} (t) \rangle$ can be  provided after selecting $f(\tau)$.
Starting from now, for tractability reasons, we assume that renewal intervals follow the Pareto's distribution
\begin{equation}
 f(\tau)=
 \left\{
 \begin{array}{ll}
  \frac{\alpha \delta^\alpha}{\tau^{\alpha+1}} & s\geqslant \delta \\
  0 & s < \delta 
 \end{array}
 \right. 
 \label{eq:pareto}
\end{equation}
with the cumulative density $F(\tau)=1-\delta^\alpha \tau^{-\alpha}$.
The Pareto density is an example of the heavy-tailed, power-law distribution \cite{nagar2016diffusion}.
The average energy 
\begin{equation}
 \langle \mathcal{E} (t) \rangle = \Lambda \int_\delta^t g_t(s) s ds=\frac{\Lambda}{\langle \tau \rangle} \int_\delta^t (1-F(s))s ds \propto t^{2-\alpha}.
 \label{eq:scaling}
\end{equation}
Consequently, for the Pareto renewal interval distribution with $1<\alpha<2$ the average energy grows as a power-law with the exponent $2-\alpha$. 
For $\alpha<1$, the average renewal interval diverges and $R(t)$ cannot be approximated by $t/\langle \tau \rangle$.
Nevertheless, we anticipate linear growth of $\langle \mathcal{E} (t) \rangle$ as it is the limiting growth in the absence of resetting.
The linear growth is already recovered for $\alpha\to 1^+$, as for $\alpha\to 1^+$ the exponent $2-\alpha$ tends to 1, see Eq.~(\ref{eq:scaling}).

\begin{figure}[htp]
 \centering
 \includegraphics[width=0.9\columnwidth]{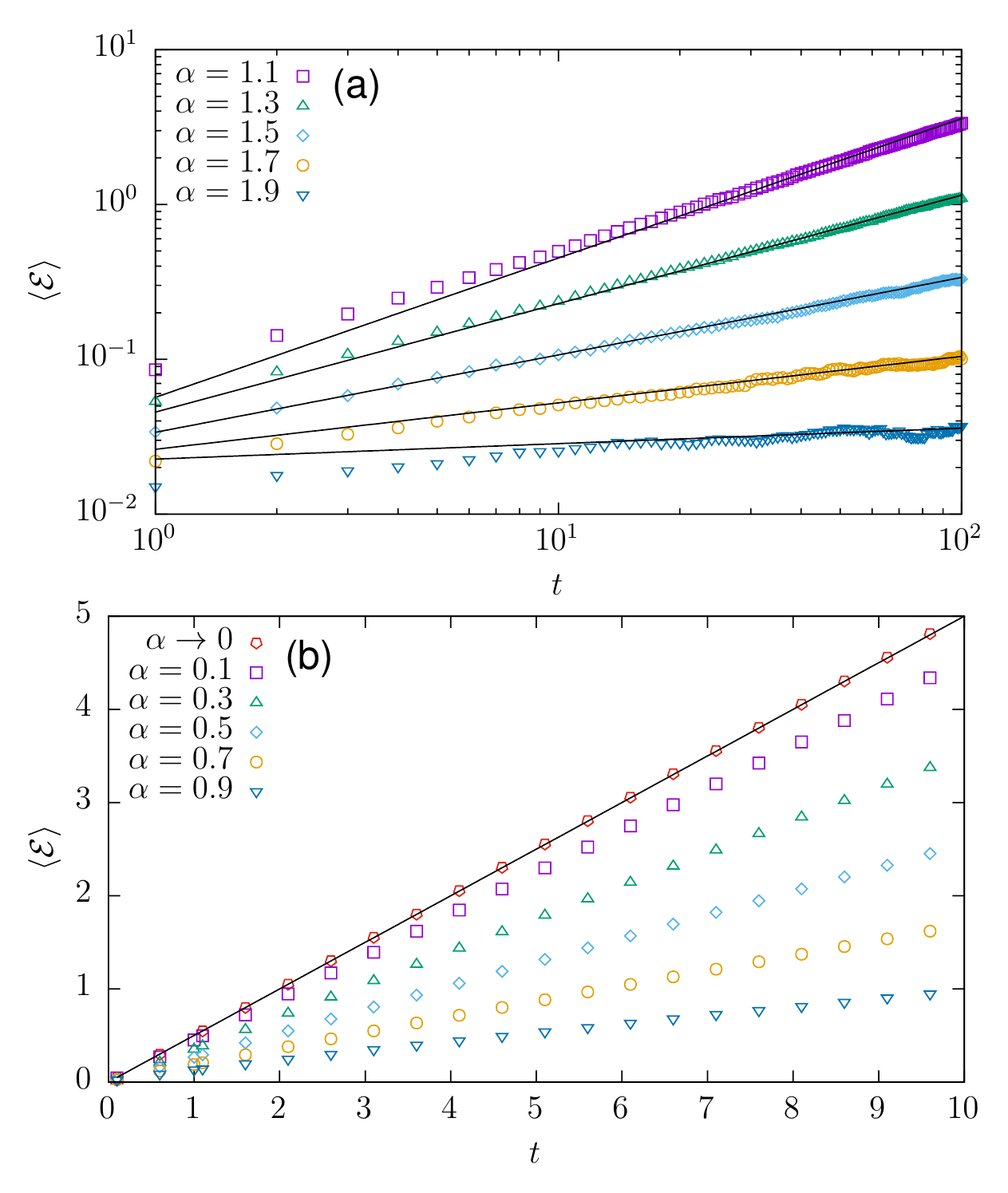}\\
   \caption{Time dependence of average energies for the parabolic potential with $(h_x,h_v)=(0,1)$ and the Pareto distribution of renewal intervals with a) $1<\alpha<2$ and b) $0<\alpha<1$.
 Various curves correspond to various values of the exponent $\alpha$. 
 Other parameters: $\delta=0.01$, $k=4$ and $m=1$. 
 Solid lines present the theoretical $t^{2-\alpha}$ scaling.
 }
 \label{fig:n2-pareto-reset}
\end{figure}

For comparison with numerical simulations, Fig~\ref{fig:n2-pareto-reset}a presents time dependence of average energy $\langle \mathcal{E}(t) \rangle$ for the parabolic potential with the Pareto distribution of renewal time intervals with $1<\alpha<2$. Various curves correspond to various exponents $\alpha$.
For $1<\alpha<2$ the Pareto distribution is characterized by the finite mean renewal time interval, but the variance of the renewal time intervals diverges.
Therefore, we are in the regime of validity of Eq.~(\ref{eq:scaling}).
The minimal renewal interval is set to $\delta=0.01$. Additional parameters are equal to $m=1$, $k=4$.
We show only results for $(h_x,h_v)=(0,1)$  as for 
$(h_x,h_v)=(1,0)$ and $(h_x,h_v)=(1,1)$ the same scaling is recorded.
Asymptotically, the average energies $\langle \mathcal{E} (t) \rangle$ scale in accordance to the predictions of Eq.~(\ref{eq:scaling}) which are depicted by solid lines.
Moreover, with decreasing $\alpha$ the quality of the $t^{2-\alpha}$ approximation improves.
For $\alpha$ tending to $1^+$ the $t^{2-\alpha}$ scaling approaches the linear scaling.
The very same scaling is recorded for average kinetic and potential energies (results not shown).
For the quartic potential, the situation is more complex because average energies do not need to grow linearly in time.
Nevertheless, for the fully traceable case of $h_x=0$, we observe the same scaling as for the parabolic potential, i.e., $t^{2-\alpha}$ (results not shown).

The case of $\alpha<1$ needs to be considered separately as for $\alpha<1$ the mean renewal time interval does not exist.
For $\alpha<1$, the linear scaling of average energy is perfectly visible in Fig.~\ref{fig:n2-pareto-reset}b.
The solid line in Fig.~\ref{fig:n2-pareto-reset}b presents the $\langle \mathcal{E} (t) \rangle=\Lambda t$ line.
In the limit of $\alpha\to 0$ there is no resetting.
For $\alpha=0.1$ the motion still can be reset but the dependence of $\langle \mathcal{E}(t) \rangle$ is very close to the one without resetting.
Change in $\alpha$, as long as $\alpha<1$, does not change the linear scaling but changes the prefactor in the scaling.

For $\alpha<1$, the ratios of average energies corresponding to Fig.~\ref{fig:n2-pareto-reset}b  asymptotically follow Eqs.~(\ref{eq:ek-partition}) and~(\ref{eq:ep-partition}) for all studied values of $h_x$ and $h_v$ (results not shown).
The situation for $\alpha>1$ is more complex.
As all types of average energies display the same scaling, we expect a kind of generalized equipartition relations with the exact shape determined by the distribution of the renewal time intervals.
For $f(\tau)$ used in Fig.~\ref{fig:n2-pareto-reset}a relations given by Eqs.~(\ref{eq:ek-partition}) and (\ref{eq:ep-partition}) are recovered for $\alpha\to 1$.
Otherwise, ratios of average energies are different.
Violations of generalized equipartition relations are due to too frequent resets.
Despite the fact that the Pareto density with $1<\alpha<2$ is characterized by the diverging variance, it still allows for frequent resetting.
The number of very short renewal time intervals can be reduced either by decreasing $\alpha$ or by increasing $\delta$.
Therefore, we have performed additional simulations with $\delta=5$, which is well above the transient period.
For $\delta=5$ we have recovered equipartition relations~(\ref{eq:ek-partition}) and~(\ref{eq:ep-partition}) (results not shown).
Here, the situation is similar to the one observed for $f(\tau)=\delta(\tau-\Delta)$ and $f(\tau)$ given by Eq.~(\ref{eq:onesidedgauss}):
too frequent resets introduce violations to equipartition relations.

\section{Summary and conclusions\label{sec:summary}}

Friction-less, noise driven, dynamics in single well potential allow for unlimited growth of average energies.
Here, we presented a possible mechanism which can assure existence of nonequilibrium stationary states.
The reintroduction of stationary states and bounding of energy is achieved by resetting velocity and position at random time instants.
If the variance of the renewal time intervals is finite, average energies are again finite and stationary states are reestablished.
Contrary to the linear damping case, these states are not of the Boltzmann--Gibbs type.
If the variance of the renewal time intervals diverges the systems still exhibit unbounded energy growth, yet slower than the limiting growth corresponding to the absence of resetting.

Using methods of statistical physics it is possible to calculate average kinetic and potential energies as a fraction of the average total energy.
Here, we show that generalized equipartition relations hold also in setups for which a stationary state does not exist, e.g., in friction-less, noise driven motions in single-well potentials. 
In order to observe generalized equipartition relations, the observation time needs to be longer than the transient period.
Consequently, in the friction-less dynamics average energies grow in unlimited manner, but after the transient period, their ratios become constant.
The very same relations hold also when both velocity and position are perturbed by noise.
Generalized equipartition relations are recovered in systems in which resets are not too frequent.
This means, in practice, that for systems with resetting for which the nonequilibrium stationary state exist, average time since the last reset has too be longer than the length of the transient period.
If despite of resetting there are no stationary states, which takes place if the variance of renewal interval diverges, generalized equipartition relations hold if short renewal intervals are excluded.
This property is related to the fact that also in the absence of resetting equipartition relations are observed after transient period.

The studied model can be generalized in numerous ways.
For example, it is possible to consider other types of resetting. 
For instance one can reset velocity or position only \cite{gupta2019stochastic}.
After such resetting system energy is reduced to kinetic or potential energies only.
Therefore, initial conditions do not correspond to $\mathcal{E}_0=0$.
On the one hand, such scenarios analogously like simultaneous resetting of position and velocity are capable of bounding energy growth.
On the other hand, these resetting scenarios are not easily accessible analytically, as energy is not fully determined by the time since the last reset but also by the value of the energy after reset.

\begin{acknowledgments}
 This research project was supported in part by the PlGrid Infrastructure.
 Computer simulations have been performed at the Academic
Computer Center Cyfronet, AGH University of Science and Technology (Krak\'ow, Poland)
under CPU grant ``DynStoch''.

\end{acknowledgments}

\appendix

\section{Evolution of average energies}
\label{sec:evolution}

From Eq.~(\ref{eq:two-noises}), i.e.,
\begin{equation}
\left\{
\begin{array}{ccl}
\frac{dx(t)}{dt} & = & v(t) + \sqrt{h_x} \xi_x(t) \\
\frac{dv(t)}{dt} & = & -\gamma   v(t) - \omega^2 x^{2n -1}(t) + \sqrt{h_v}\xi_v(t)
\end{array}
\right.
\label{eq:two-noises-app}
\end{equation}
it is possible to derive equations describing time evolution of average kinetic $\langle \ek \rangle$ and potential  $\langle \ep \rangle$ energies.
Starting from $\ep=k \frac{x^{2n}}{2n}$ we have
\begin{eqnarray}
 d\ep & = &  \frac{d\ep}{dx} dx + \frac{1}{2}\frac{d^2\ep}{dx^2} (dx)^2+\dots \\ \nonumber
 & =& kx^{2n-1}dx+\frac{1}{2}k(2n-1)x^{2n-2}(dx)^2
\end{eqnarray}
with 
\begin{equation}
dx=vdt+\sqrt{h_x}dW_x ,    
\label{eq:dx}
\end{equation}
where $dW_x$ stands for increments of the Wiener $W_x$ process (Brownian motion).
Analogously for $\ek=\frac{1}{2}m v^2$ we have
\begin{eqnarray}
 d\ek & = &  \frac{d\ek}{dv} dv + \frac{1}{2}\frac{d^2\ek}{dv^2} (dv)^2+\dots \\ \nonumber
 & =& mvdv+\frac{1}{2}m(dv)^2
\end{eqnarray}
with 
\begin{equation}
dv=-\gamma v dt -\omega^2 x^{2n-1} dt +\sqrt{h_v} dW_v, 
\label{eq:dv}
\end{equation}
where $dW_v$ represents increments of the Wiener process $W_v$.
We assume that both processes $W_x$ and $W_v$ are independent, i.e., $\langle \xi_x(t)\xi_v(s) \rangle=0$.
Keeping terms which are at most linear in $dt$ we have
\begin{equation}
d \ep= k x^{2n-1} v dt + \sqrt{h_x} k x^{2n-1} dW_x+\frac{1}{2}h_xk(2n-1)x^{2n-2}dt
\end{equation}
and
\begin{equation}
 d\ek=-m\gamma v^2dt-m\omega^2 x^{2n-1}v dt+m \sqrt{h_v}v dW_v+\frac{1}{2}mh_vdt.
\end{equation}
From above equations, we obtain following formulas for time derivatives of average energies

\begin{equation}
 \frac{d}{dt}\langle \ep (t) \rangle= k \langle x^{2n-1} v \rangle + \frac{1}{2}h_x k(2n-1)\langle x^{2n-2} \rangle
\label{eq:epgrowthrate-app}
\end{equation}
and

\begin{equation}
 \frac{d}{dt}\langle \ek (t) \rangle= -m\gamma \langle v^2 \rangle -m\omega^2 \langle x^{2n-1} v \rangle +\frac{1}{2}mh_v,
 \label{eq:ekgrowthrate-app}
\end{equation}
since the correlators  $\langle  x^{2n-1} dW_x \rangle$ and $ \langle v dW_v \rangle$ vanish \cite{tome2015stochastic}.
Change of total energy is described by
\begin{eqnarray}
\label{eq:growthrate-app}
 \frac{d}{dt} \langle \mathcal{E} (t) \rangle & = & \frac{d}{dt} \langle \ep (t) + \ek (t) \rangle \\ \nonumber
 & = & -m\gamma \langle v^2\rangle + \frac{1}{2}h_x k(2n-1)\langle x^{2n-2}\rangle + \frac{1}{2}mh_v. 
\end{eqnarray}
Alternatively to It\^o lemma one can use the method described in Chapter~3 of \cite{tome2015stochastic}, which gives the same results.

\def\url#1{}


\end{document}